\documentstyle[preprint,prc,aps]{revtex}
\begin{document}
\preprint{RUB-MEP-93/94}
\draft
\title{Sensitivity of pp-Bremsstrahlung to Meson-Exchange Currents}
\author{J. A. Eden and M. F. Gari}
\address{ Institut f\"ur Theoretische Physik\\
	  Ruhr Universit\"at Bochum, D-44780 Bochum, Germany\\
	  {\rm email: jamie@deuteron.tp2.ruhr-uni-bochum.de}}
\date{30 August 1994; Revised 10 January 1995}
\maketitle
\tighten
\begin{abstract}
We present pp-bremsstrahlung cross section calculations at the
$\pi$-production threshold to search for kinematics that best differentiate
model-dependent descriptions of the NN $t$-matrix and accentuate the
contributions of isoscalar meson-exchange currents.
Existing optimization procedures are shown to be completely unreliable
due to the neglect of meson-exchange currents and phase space variations.
We show that phase-space hides the most important differences in model
$t$-matrices,
but that improved data will be critical to further investigations of exchange
currents.
\end{abstract}
\pacs{PACS:13.75.Cs, 25.20.Lj}
%
The dominant processes in pp-bremsstrahlung at energies near the
$\pi$-production threshold are the nucleon-pole contributions shown in
Fig.~1(a-b).
It has long been thought \cite{AS49} that this `nucleon-pole dominance'
would allow for the investigation of the off-shell behavior of various
NN-interaction models, nearly all of which are fitted to the on-shell
NN-scattering data. However, within the assumption of nucleon-pole dominance,
gauge invariance and the soft-photon theorem \cite{LO58,AD66} establish that
the first two terms in the expansion of the bremsstrahlung amplitude in powers
of the photon energy are completely specified by on-shell scattering
\cite{NY68,FE67}.

To obtain new information from bremsstrahlung that is not already given by
on-shell NN-scattering data, the kinematics of a recent experiment at TRIUMF
\cite{TRIUMF} were optimized \cite{HF79} to obtain large photon energies and
highly off-shell nucleon-pole contributions.
In earlier work \cite{JE93} we compared this data from TRIUMF with relativistic
meson-baryon model calculations that include only the processes of
Fig.~1(a-c), but we obtained results that share the discrepancy with experiment
reported elsewhere \cite{HF86,VB91,VH91,VH92,HF93a,HF93b,KA93}.
We have since repeated these calculations using wave functions obtained from
the Paris \cite{Paris},
Nijmegen \cite{Nijmegen}, Bonn B \cite{Bonn} and RuhrPot \cite{RuhrPot}
NN-interactions, but we obtain results that are essentially all the same and
continue
to show a collective discrepancy with experiment. We realize that in all but
the most forward scattering angle data, such discrepancies cannot be explained
by
any uncertainty in experimental absolute scale.

There are two important points here. The first is that,
on the assumption of nucleon-pole dominance with maximum photon energies,
the TRIUMF experiment was optimized to probe differences in model-dependent
off-shell t-matrices, but the similarity of the model results indicates that it
failed to do so. In the following we will show that this result could have been
predicted from simple phase space considerations.
The second point is that the collective discrepancy between the experiment and
the impulse approximation calculations indicates that large contributions from
meson-exchange currents may be present.

We attempt here a more rigorous procedure for selecting pp-bremsstrahlung
kinematics that might differentiate the off-shell behavior of NN-interaction
models. We also identify kinematics which enhance the contributions of the
dominant isoscalar meson-exchange currents - a task which is clearly impossible
within the universally accepted procedure of optimizing experiments
according to nucleon-pole dominance.

Our procedure for calculating the impulse contributions of Figs~1(a-c) has been
described in ref~\cite{JE93}. We add to this the perturbative
$\rho\pi\gamma$, $\omega\pi\gamma$, $\rho\eta\gamma$, $\omega\eta\gamma$
and N$\Delta\gamma(\pi,\rho)$ exchange currents currents shown in Fig~1(d-e),
all of which are easily calculated in the usual way \cite{CH71,DR89,JT89,RW79}
with,
\begin{eqnarray}
{\cal L}_{{\rm PV}\gamma} &=& -{e_p g_{_{\rm PV}\gamma}\over 2m_{_{\rm V}} }
 \epsilon^{\mu\nu\sigma\tau} F_{\mu\nu} \vec{\psi}_{\sigma}^{\,^{\rm
V}}\vec{\psi}_{\tau}^{\,^{\rm P}},
\cr
{\cal L}_{{\rm N}\Delta\pi} &=& -{g_{{\rm N}\Delta\pi}\over 2m}
 \bar{\psi}^\mu \vec{\tau}_{{\rm N}\Delta} \psi \partial_\mu \vec{\pi} + {\rm
h.c.}
\cr
{\cal L}_{{\rm N}\Delta\rho} &=& - i {g_{{\rm N}\Delta\rho}\over 2m}
 \bar{\psi}^\mu \gamma^5 \gamma^\nu \vec{\tau}_{{\rm N}\Delta} \psi
\vec{\rho}_{\mu\nu} + {\rm h.c.}
\cr
{\cal L}_{{\rm N}\Delta\gamma} &=& -i {e_p\mu_{{\rm N}\Delta}\over 2m}
 \bar{\psi}^\mu \gamma^5 \gamma^\nu \vec{\tau}_{{\rm N}\Delta}^{\,3} \psi
F_{\mu\nu} + {\rm h.c.}
\end{eqnarray}
where V=$\rho,\omega$ and P=$\pi,\eta$.
In the present work we neglect contributions involving negative-frequency
particles.
To fully specify the model, we fix the coupling constants and $\Delta$-decay
width to their experimental \cite{PDG} values of
$g_{\rho\pi\gamma}$=0.55,
$g_{\omega\pi\gamma}$=2.03,
$g_{\rho\eta\gamma}$=1.39,
$g_{\omega\eta\gamma}$=0.31,
$g_{{\rm N}\Delta\pi}$=28.85, and $\Gamma_{\Delta}$=115~MeV.
For the remaining experimentally unknown quantities, we adopt the familiar
SU(6) results
$g_{{\rm N}\Delta\rho}$=$g_{{\rm NN}\rho}(1+\kappa_\rho)g_{{\rm
N}\Delta\pi}/g_{{\rm NN}\pi}$,
$\mu_{{\rm N}\Delta}$ = ${1\over 2}(1+\kappa^{{\rm iv}}) g_{{\rm
N}\Delta\pi}/g_{{\rm NN}\pi}$.
The NN-meson coupling constants and form factors are taken from
the RuhrPot\cite{RuhrPot} interaction to ensure that the
meson-exchange currents, the strong interaction potential and the strong
form factors \cite{SD90,JF92} are consistently defined.
For the real photon, the Dirac and Pauli electromagnetic form factors
obviously reduce to their normalization values, and
since the present work is confined to low $Q^2$, we
approximate $F_{{\rm N}\Delta\alpha}$ = $F_{{\rm NN}\alpha}$ for
$\alpha$=$\pi,\rho$.
No form of soft photon approximation is adopted at any stage.

There are at least three fundamental differences between our exchange current
calculations and the NN$\rightarrow$N$\Delta$ $t$-matrix calculation reported
elsewhere.
First, like \cite{MJ94}, we include
the $\rho\pi\gamma$, $\omega\pi\gamma$, $\rho\eta\gamma$ and $\omega\eta\gamma$
exchange currents that have been neglected in \cite{JO94}.
Although the $\Delta$ contributions are certainly the largest of the exchange
currents,
the vector-meson decay contributions are non-negligible when interferences are
correctly included.
Second, our perturbative treatment of the N$\Delta\gamma(\pi,\rho)$ exchange
currents
allows us to retain the meson recoil contributions that cannot be included in
any
$t_{\Delta{\rm N}}$-matrix expansion.
As is well known, the recoil and wave function re-orthonormalization
contributions cancel in the static limit for NN$\rightarrow$NN interactions
\cite{GA76},
so that their combined contributions can reasonably be ignored to permit use of
the
ladder-approximation in calculating of the corresponding $t$-matrix.
However, no such cancellation occurs in the NN$\rightarrow\Delta$N interaction,
so that
the recoil and wave function re-orthonormalization terms are neglected without
justification
in refs~\cite{MJ94,JO94}.
Third, (and relevant to the second point) we have not fitted the isobar
contributions to
experimental data and no unphysical free parameters are required to avoid
double-counting
of N$\Delta$ intermediate states. As such, the isobar mass, coupling constants
and form
factors we use are not effective fit parameters and can be unambiguously
identified
with the physical nucleon resonance $\Delta(1232)$.
In addition, we note that the impulse approximation rescatter amplitudes
represent a correction
of less than 15\% to the dominant nucleon-pole contributions, and that this
provides a crude
measure of the accuracy of the perturbative exchange current amplitudes we
include here.
In view of the differences between the present approach and that reported in
ref~\cite{JO94}
the numerical results of the two calculations are fairly similar,
at least at the kinematics where comparison is possible.

For the kinematically complete bremsstrahlung reaction
$p_1+p_2\rightarrow p_3+p_4+k$, data is usually presented as a function of
the photon emission angle $\theta_\gamma$ for given proton scattering angles
$\theta_3$ and $\theta_4$ and non-coplanarity
$\Phi={1\over 2}(\pi+\phi_3-\phi_4)$ \cite{DM68}. All quantities are
given in the Lab frame $(\vec{p}_2=0)$, and we denote $\theta_3$ ($\theta_4$)
as the scattering angle for the proton that is emitted on the same (opposite)
side of the beam as the photon. Since we describe the hadronic wave functions
non-perturbatively with NN t-matrices, we avoid the approximation of ignoring
inelasticities by confining our present investigations to energies below the
$\pi$-production threshold.

In the left-hand column of Fig~2, we present the ratio of pp-bremsstrahlung
cross sections at E$_{\rm lab}$=280 MeV including the nucleon-pole and
rescatter
contributions of Fig.~1(a-c), and using wave functions obtained from the
RuhrPot
\cite{RuhrPot} and Bonn B \cite{Bonn} potentials.
In the middle column, we present similar cross section ratios at corresponding
kinematics and use wave functions obtained from the RuhrPot model with
the nucleon-pole and rescatter contributions of Fig~.1(a-c), and which
either retain (IA+Mexc) or neglect (IA) the meson-exchange currents of
Fig~.1(e-f). The phase space available to these reaction geometries is shown in
the right-hand column.

In Figs.~2a(1-3) we consider coplanar geometries with $\theta_3$=$\theta_4$.
Figs.~2a(1) shows the potential models have only weak and structureless
differences at nearly all geometries, whereas Figs.~2a(2) shows that the
isovector exchange currents
are large near $\theta_3=\theta_4\sim 20^\circ$.
We realize that recommendations \cite{HF79,VB91} based nucleon-pole dominance
with maximum photon energies call for increased emphasis on the extreme
forward-angle geometries.  Although these geometries succeed in maximizing the
differences between the potential-model calculations, Figs.~2a(3) shows that
only the regions of $\theta_\gamma\sim 0^\circ$ and 180$^\circ$ have
significant
phase space, and here Figs.~2a(2) shows that such regions take non-negligible
contributions from meson-exchange currents.

Upon the recommendations of nucleon-pole dominance, an experiment was recently
reported \cite{HOM92a} at E$_{\rm lab}\sim294$ MeV and
$\theta_3=\theta_4=6.0^\circ\pm1.2^\circ$. However,
the data suffers from very large phase space variations over the detector
acceptance angles and simply reflects the peculiarities of various experimental
efficiencies.  As such, a meaningfully comparison with theory is impossible.
The fact that this region is of limited interest has only recently been
appreciated in a phenomenological analysis \cite{HOM92b}, although this lead to
some conclusions that we cannot corroborate. Similar difficulties in future
experiments
can be avoided by adopting an analysis like that that shown in Fig~2.

In Figs.~2b(1-3) and Figs.~2c(1-3) we consider coplanar scattering with
$\theta_3\neq\theta_4$. Again we observe large contributions from
the isovector exchange currents, but structured differences in the
potential-model calculations occur only in regions suppressed by phase space.

Finally, Figs.~2d differ from Figs.~2a in that we fix $\theta_3=\theta_4=20$
and
scan the kinematically permissible non-coplanar geometries. From the outset,
such geometries are undesirable since the dynamic range of $\theta_\gamma$ is
suppressed by momentum conservation and enormous phase space variations
will dominate any data which includes events near the kinematic limit of
non-coplanarity (for example, the data of ref \cite{HOM92a}).
In addition, we realize that a dramatic improvement in the data will be
required before non-coplanar analyzing powers \cite{KA93} will be of interest.
This shows that increasing the non-coplanarity offers no immediate gains to
off-set some inherent disadvantages, at least at $\theta_3\sim\theta_4\sim 20$.

Fig~2 establishes that the exchange currents are manifest in pp-bremsstrahlung
in regions where phase space is large and smooth. We recommend an increased
experimental
emphasis on coplanar reaction geometries involving larger proton scattering
angles.
To compare our results with others \cite{MJ94,JO94}, in Fig~3. we provide a
conventional representation of the exchange current effects and observe a
qualitative
similarity amoungst all of the recent exchange current calculation results.

We conclude that pp-bremsstrahlung data at energies near the $\pi$-production
threshold is unlikely to distinguish the differences between potential models
because phase space suppresses the cross section in precisely the regions where
the off-shell $t$-matrix elements are most different. This has not been
recognized in the past analyses, all of which are essentially based on
nucleon-pole dominance and, as such, neglect the critical roles of phase space
and meson-exchange currents.
However, a modest reduction in the experimental
uncertainty of the coplanar data near 0$\leq\theta_3\leq 25$ and
$\theta_4\sim 20$ would provide extremely useful information for the further
investigation of relativistic- and resonance width-effects in T=1 isoscalar
exchange currents. The current emphasis on highly non-coplanar geometries
should probably be reconsidered.

\acknowledgements
This work is supported by COSY-KFA J\"ulich (41140512).


\begin{figure}
\caption{Dominant contributions to pp-bremsstrahlung observables:
(a-b) nucleon-pole and (c) rescatter contributions to the impulse approximation
(IA) current, and (d) N$\Delta\gamma(\pi,\rho)$ and
(e) $\rho\pi\gamma$, $\omega\pi\gamma$, $\rho\eta\gamma$ and $\omega\eta\gamma$
meson-exchange currents (Mexc). All such contributions are included in the
exchange-current calculations of the present work.}
\end{figure}
%
\begin{figure}
\caption{ pp-bremsstrahlung calculation results at E$_{\rm lab}$=280 MeV for
proton
scattering angles of $\theta_3$ and $\theta_4$,
photon emission angle $0\leq\theta_\gamma\leq 180^\circ$ and non-coplanarity
$\Phi$.
In Fig~2(a-d)(1) we explore model dependence by plotting cross sections ratios
for the RuhrPot and Bonn models including the processes of Fig.~1(a-c).
In Fig~2(a-d)(2) we explore exchange currents by plotting cross sections ratios
for the RuhrPot model including the processes (IA) of Fig~1(a-c) and
(IA+Mexc) of Fig~1(a-e).
In Fig~2(a-d)(3) we show the phase space for the reaction geometry.
The spikes in Fig~d(3) simply reflect the plot grid points in the region of
the square-root singularity at the kinematic limit of non-coplanarity [30].
Phase space is either small or structured where the models are most different,
but is
smooth and large where the exchange currents are important.}
\end{figure}
%
\begin{figure}
\caption{ Coplanar pp-bremsstrahlung calculation results at E$_{\rm lab}$=280
MeV.
We find the $\rho\pi\gamma$, $\omega\pi\gamma$, $\rho\eta\gamma$,
$\omega\eta\gamma$
and N$\Delta\gamma(\pi,\rho)$ exchange currents enhance the cross section for
$\theta_3$=28.0 and $\theta_4$=27.8 near $\theta_\gamma$=75\% by about 50\%.
Given that the $\rho\pi\gamma$, $\omega\pi\gamma$, $\rho\eta\gamma$ and
$\omega\eta\gamma$
exchange currents are ignored in [28], we realize that results are qualitately
similar
 - the corresponding enhancement there being $\sim$ 45\%.
This geometry is a good candidate for better measurement since the cross
section
is large, phase space is smooth, and the contributions of the exchange currents
are clearly visible.
}
\end{figure}
\end{document}